\documentclass[pra,aps,twocolumn,showpacs,superscriptaddress,dvipsnames]{revtex4}

\usepackage{amsmath,amssymb,bm,amsthm,bbold}
\usepackage{graphics,epsfig}
\usepackage[colorlinks=true,citecolor=blue,linkcolor=blue,urlcolor=blue]{hyperref}
\usepackage[T1]{fontenc}
\usepackage{mathptmx}

\def\tr{\mbox{Tr}}
\def\be{\begin{equation}}
\def\ee{\end{equation}}

\newcommand{\la}{\left\langle}
\newcommand{\ra}{\right\rangle}
\newcommand{\zero}{\left|0\ra}
\newcommand{\one}{\left|1\ra}
\newcommand{\ket}[1]{\left| #1 \ra}
\newcommand{\bra}[1]{\la #1 \right|}
\newcommand{\vac}{\ket{\varnothing}}

\newcommand{\modsq}[1]{\left| #1 \right|^2}

\newcommand{\acrea}[1]{\hat{a}_#1^\dagger}
\newcommand{\lb}[1]{\log_2\left( #1 \right)}

\begin{document}

\title{Information gap for classical and quantum communication in a Schwarzschild spacetime}

\author{Dominic Hosler}\email{dominichosler@physics.org}
\affiliation{Department of Physics and Astronomy, University of Sheffield, Hicks building, Hounsfield road, Sheffield, S3 7RH, United Kingdom}

\author{Carsten van de Bruck}\email{C.vandeBruck@sheffield.ac.uk}
\affiliation{School of Mathematics and Statistics, University of Sheffield, Hicks building, Hounsfield road, Sheffield, S3 7RH, United Kingdom}

\author{Pieter Kok}\email{p.kok@sheffield.ac.uk}
\affiliation{Department of Physics and Astronomy, University of Sheffield, Hicks building, Hounsfield road, Sheffield, S3 7RH, United Kingdom}

\begin{abstract}
\noindent
Communication between a free-falling observer and an observer hovering above the Schwarzschild horizon of a black hole suffers from Unruh-Hawking noise, which degrades communication channels. Ignoring time dilation, which affects all channels equally, we show that for bosonic communication using single and dual rail encoding the classical channel capacity reaches a finite value and the quantum coherent information tends to zero. We conclude that classical correlations still exist at infinite acceleration, whereas the quantum coherence is fully removed.
\end{abstract}


\date{\today}
\maketitle

\section{Introduction}
\noindent
It has been suggested that physics can and should be formulated in the language of information theory, where all interactions are viewed as information transfer between agents \cite{Wheeler1989}. The channel capacity represents the maximum amount of information that can be transferred in such interactions. When the channel capacity between two systems, $A$ and $B$,  drops to zero, $B$ cannot learn anything about the state of $A$. Consequently, $A$ cannot have any effect on $B$. This is a stronger condition on the range of influence than mere locality, in which $A$ and $B$ have to be outside each other's light cones to prohibit interactions between them. This fundamental link between information channels and causality can further our knowledge of the structure of a quantum theory of gravity. When $A$ and $B$ are quantum mechanical systems, the communication channels that can be defined between them are described by quantum information theory \cite{Nielsen2000,Desurvire2009}. Quantum communication channels are generally different from the classical channels, because there are more ways we can communicate with quantum systems than there are with classical systems \cite{Shannon1949,Bennett1992,Schumacher1997,Abeyesinghe2009,Lupo2010}. Their channel capacities therefore require different definitions \cite{Horodecki2005,Horodecki2006}.

Quantum channel capacities have been studied extensively in non-relativistic settings, but to date relatively little work has been done on relativistic quantum information channels. In relativistic quantum information theory \cite{Peres2004a,Terno2005} the structure of spacetime affects the ways in which information can be sent from $A$ to $B$ \cite{Alsing2003,Kok2006,Cliche2010,Cliche2010a}. The presence of horizons will introduce noise in the form of Unruh-Hawking radiation \cite{Fuentes-Schuller2005,Martin-Martinez2010b,Mathur2009,Smolin2011,Pan2008}, and generally reduce the quantum channel capacity \cite{Bradler2010}. Trade-off capacities have been studied, where the rate of classical and quantum communication is traded off with rate of entanglement consumption \cite{Bradler2010b,Jochym-OConnor2011a,Wilde2011c}.

In this paper we study the classical channel capacity and quantum coherent information, which allows us to compare their behaviour in the situation where Alice and Rob communicate near a black hole. In Section \ref{sec:setup} we present the particular communication setup we consider in this paper. In Section \ref{sec:modes} we describe the transformation from Rindler modes to Unruh modes, and in Section \ref{sec:channels} we calculate the classical channel capacities and coherent information. Finally, we present our conclusions in Section \ref{sec:conclusions}.

\section{Communication Setup}\label{sec:setup}
\noindent
Imagine the situation where Alice is an inertial observer free-falling into a Schwarzschild black hole, while Rob hovers at a fixed distance from the horizon. Alice wants to send a message to Rob. As shown in \cite{Martin-Martinez2010a}, for the case where Rob is near the horizon of a large black hole, this situation is equivalent to a constantly accelerating observer in Minkowski space. We compare classical and quantum communication in the setting of two physical encoding methods using bosonic fields. The classical case is where Alice creates two correlated bits and transfers one to Rob. In the quantum case Alice creates an entangled pair of qubits and transfers one to Rob. Both protocols are analysed using two encoding methods, namely \emph{single rail} and \emph{dual rail}. Single rail represents a $\zero$ and $\one$ by the absence and presence of an excitation (photon), respectively. In the dual rail encoding there is always exactly one excitation that can exist in a superposition of two modes. We
calculate the distinguishability of the qubit states, as well as the channel capacity and coherent information of the various channels. We concentrate on the effect of the Unruh-Hawking noise, and ignore the reduction of the channels due to the gravitational redshift, since this affects all channels in equal measure.
These channels will inform us of the way quantum information is affected by general relativistic situations.

To quantify information, we use the Von Neumann entropy $S(\rho) = -\tr[\rho \lb{\rho}] = -\sum_i \lambda_i \lb{\lambda_i}$, where $\lambda_i$ are the eigenvalues of the density matrix $\rho$. This reduces to the Shannon entropy when the $\lambda_i$ are elements of a classical probability distribution. We measure the amount of classical information that is shared between Alice and Rob by the {\em mutual information} $S(\rho_A;\rho_R) \equiv S(\rho_A)+S(\rho_R)-S(\rho_{AR})$, where $\rho_{AR}$ is the state of the joint system, and $\rho_A = \tr_R(\rho_{AR})$ and $\rho_R = \tr_A(\rho_{AR})$ are the reduced density operators for Alice and Rob's subsystems \cite{Desurvire2009}. Quantum mechanically, the mutual information is a measure of the correlations between Alice and Rob, which can manifest itself as entanglement: a system can have finite mutual information even if the state $\rho_{AR}$ is pure [$S(\rho_{AR})=0$]. In addition, for quantum communication we want to know the amount of extra information that is
required for Rob to fully specify Alice's state. This can be negative, which is interpreted as the amount of information that may be sent at a later time using the correlations of the channel. This quantity is the {\em conditional entropy} $S(\rho_A | \rho_R) \equiv S(\rho_{AR})-S(\rho_R)$ \cite{Horodecki2005}. The classical channel capacity is the maximum of the mutual information, and is interpreted as the classical information shared per use of the channel \cite{Shannon1949}. The quantum coherent information is the negative conditional entropy, and it is interpreted as the amount of information per use of the channel that Alice and Rob can communicate in the future, given unlimited classical communication \cite{Horodecki2005}.

Alice prepares a qubit in the state $\rho$ and correlates it with a photonic qubit via $\ket{0}_A\to\ket{0}_A\ket{0}_p$ and $\ket{1}_A\to\ket{1}_A\ket{1}_p$. The photon is sent to Rob and the joint system is in the state $\rho_{AR}$. Rob measures his local mode with an ideal detector. We consider the various information channels that are established this way. Particle number is not conserved due to Unruh-Hawking noise, and Alice and Rob will have to choose an error correction scheme and measurement basis that optimises the amount of information sent through the channel. Here we will assume that this is possible and consider only the channels themselves.

\section{Mode transformations}\label{sec:modes}
\noindent
Alice describes the field using Minkowski modes, which form a complete orthonormal set in Minkowski space-time. Since Rob observes a horizon, we need to separate his description of the field into two causally disconnected parts, called region I (inhabited by Rob) and region IV (inhabited by ``anti-Rob''). Rob and anti-Rob define so-called Rindler modes in their respective regions, which together cover all space-time. The Rindler modes approximate the Schwarzschild modes required to describe field modes in a Schwarzschild metric \cite{Martin-Martinez2010b}. We can therefore draw conclusions about Schwarzschild black holes in the appropriate limit from studying the mathematically simpler Rindler space-time.

Alice has the freedom to create excitations in any accessible mode. Hence, we choose Alice's modes as superpositions of different frequencies such that each of Alice's modes maps to single-frequency Rindler modes $\omega$, forming a Minkowski packet $\hat{a}^\dagger_P$.
We then transform it to the creation operators of Unruh modes, $\hat{A}^\dagger_L$ and $\hat{A}^\dagger_R$ for left and right wedges respectively, using the most general transformation $\hat{a}^\dagger_P = q_L \hat{A}^\dagger_L + q_R \hat{A}^\dagger_R$, where $q_L$ and $q_R$ are complex numbers such that $\modsq{q_L}+\modsq{q_R}=1$ \cite{Bruschi2010}.
This means that the Minkowski packets can relate to a superposition of Unruh modes, in the left and right wedges. The single mode approximation is where we set $q_R=1$ and $q_L=0$, breaking the symmetry between the left and right Unruh wedge, which correspond to the Rindler wedges I and IV, respectively.  Using the Unruh modes allows us to maximise the correlations between Alice and Rob. Since we are interested in communication channels we maximise the mutual information and the coherent information. This maximisation forces the choice of $q_R$ and $q_L$ to match the single mode approximation.

The transformation between the Unruh and Rindler modes is given by the two-mode squeezing operator $U = \exp[ i r ( \hat{a}_I \hat{a}_{IV} + \hat{a}_I^\dagger \hat{a}_{IV}^\dagger )  ]$, with $r$ a real squeezing parameter and $\hbar=c=G=1$ \cite{Unruh1976}.
This parameter is related to Rob's proper acceleration $a$ via $\tanh r = \exp(-\omega\pi/a)$ and can be used to approximate the proper acceleration of Rob at a distance $R$ from the black hole with mass $M$ and Schwarzschild radius $R_S$ as $a^{-1} = 4M\sqrt{1-R_S/R}$ \cite{Martin-Martinez2010a}.
The transformation mixes creation and annihilation operators and does not preserve photon number. However, unitarity ensures that the transformation preserves information globally.

Using the Baker-Campbell-Hausdorff theorem for optical modes \cite{Kok2010} we can write the effect of the transformation on the Minkowski vacuum and single photon state as
\begin{align}
  \label{eq:vacuumtransform}
 \hat{U} \vac & = \frac{1}{\cosh r} \sum_{n=0}^\infty \tanh^n r \ket{n_I,n_{IV}}, \\
  \label{eq:particletransform}
\hat{U} \acrea{I} \vac & = \frac{1}{\cosh^2 r} \sum_{n=0}^\infty \tanh^n r\; \sqrt{n+1} \ket{(n+1)_I,n_{IV}}.
\end{align}
This operation creates entanglement between Rindler modes in region I (Rob) and in region IV (Anti-Rob). Rob is causally disconnected from region IV and has no access to information there so we must trace out all region IV modes. This results in an effective non-unitary transformation and local information loss. In the Rindler vacuum $R$ the density operator elements of the Minkowski modes $M$ become
\begin{align}
 \label{eq:transformelementjk}
 \ket{j}_M\bra{k} \to& \sum_{n=0}^\infty \frac{\tanh^{2n}r}{\cosh^{(2+j+k)}r}\left(n+1\right)^{\frac12(j+k)}\ket{n+j}_R\bra{n+k}\, ,
\end{align}
where $j,k\in\{0,1\}$.
We consider two logical states prepared by Alice, namely the classical state $\rho = |\alpha|^2 \ket{0}\bra{0} + |\beta|^2 \ket{1}\bra{1}$ and the  qubit state $\ket{\psi} = \alpha\ket{0} + \beta\ket{1}$. The former leads to strictly classical correlations between Alice and Rob, while the latter generally leads to entanglement. We will analyse the mutual information and conditional entropy that these states will give rise to.

\begin{figure}[t]
  \begin{center}
      \includegraphics[width=8.5cm]{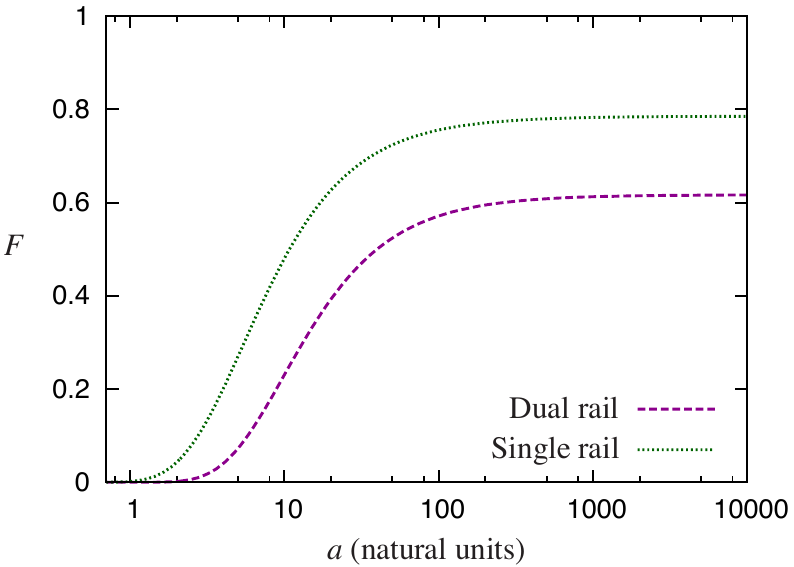}
  \caption[Fidelity between recieved zero and one]{Fidelity between the logical zero and one states received by Rob as a function of the acceleration $a$ for single and dual rail qubits.}  \label{fig:fidelity-of-zero-vs-one}
  \end{center}
\end{figure}

To quantify how the channel reduces the distinguishability between the logical states sent by Alice, we use the probability of mistaking $\rho_0 = \ket{0}_M\bra{0}$ for $\rho_1=\ket{1}_M\bra{1}$ when performing any measurement, given by the fidelity $F = \tr [\sqrt{\sqrt{\rho_0}\rho_1\sqrt{\rho_0}}]^2$ \cite{Jozsa1994}.
This is calculated using the states to which Rob has access, so when Alice sends a logical zero his state is given by Eq.~(\ref{eq:transformelementjk}) with $j=k=0$, and when Alice sends a logical one his state is given by $j=k=1$. These states are diagonal in the same basis, and the fidelity reduces to
\begin{equation}
 \label{eq:overlapbetween0and1}
F = \tr \left[\sqrt{\rho_0\rho_1}\right]^2 = \left(\sum_{n=1}^\infty \frac{\tanh^{2n-1}r}{\cosh^3r}\sqrt{n}\right)^2\, ,
\end{equation}
shown in Fig.~\ref{fig:fidelity-of-zero-vs-one} for both dual rail and single rail encoding. For small $a$ the system approaches a perfect channel, indicated by a near-zero fidelity between the logical states. As $a$ increases, the probability of Rob getting an incorrect measurement result increases, reducing the channel channel capacity. Note that when $a\to\infty$ we find $F\to F_0<1$. This is due to the persistent difference of a single photon in the two Rindler states. Consequently, the classical channel capacity should never drop to zero.

\section{Communication channels}\label{sec:channels}
\begin{figure}[th]
  \begin{center}
      \includegraphics[width=8.5cm]{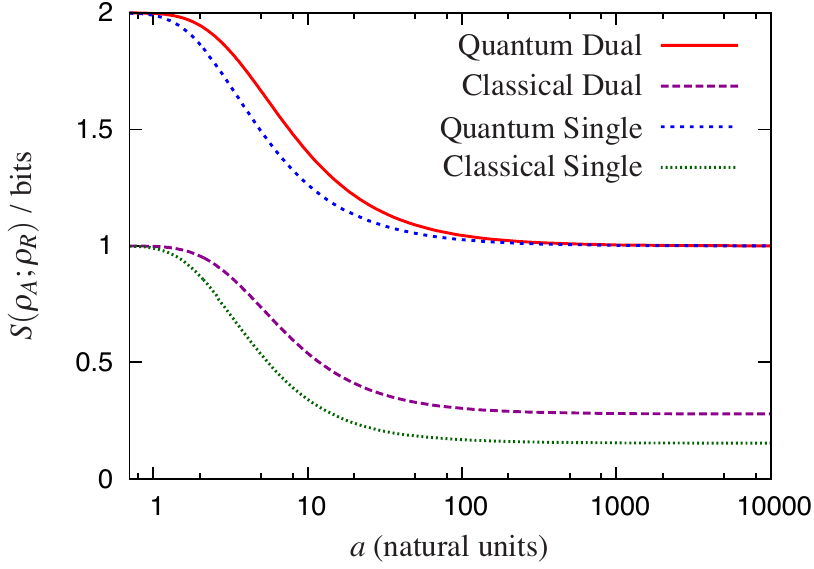}
  \caption[Mutual Information]{Mutual information as a function of the  acceleration $a$ when $\modsq{\alpha} = \modsq{\beta} =\frac12$.}
  \label{fig:ME-compare-single-dual-mixed-full}
  \end{center}
\end{figure}

\noindent
In this section we calculate and compare the channel capacities and coherent information of single and dual rail classical and quantum communication. Dual rail communication is physically symmetrical in the logical zero and logical one states, which means that the maximum mutual information is obtained when $\modsq{\alpha} = \modsq{\beta} =\frac12$. For single rail encoding the situation is slightly more complicated so we give the full expressions here.
\begin{widetext}
We calculate the mutual information for the classical case $\rho$ in the single rail encoding as
\begin{align}\label{eq:MIclassicalsingle}
S\left(\rho_A;\rho_R\right) = & -\modsq{\alpha} \lb{\modsq{\alpha}} - \modsq{\beta}\lb{\modsq{\beta}} \\
                             & - \sum_{n=0}^\infty \left[ \frac{\modsq{\alpha}}{\cosh^2r} \tanh^{2n}r \lb{1+\frac{n \modsq{\beta}}{\modsq{\alpha}\sinh^2r}}
                                + \frac{n \modsq{\beta}}{\cosh^2r\sinh^2r}\tanh^{2n}r \lb{1+\frac{\modsq{\alpha}\sinh^2r}{n \modsq{\beta}}} \right] ,\nonumber
\end{align}
The mutual information for the classical case, in the dual rail encoding, having substituted $\modsq{\alpha} = \modsq{\beta} =\frac12$ is calculated as
\begin{align}
  \label{eq:MIclassicaldual}
S\left(\rho_A;\rho_R\right)= 1 - \sum_{p=0}^\infty  \frac{\tanh^{2p}r}{\cosh^6r} \sum_{q=0}^p (q+1)\lb{1+\frac{p-q}{q+1}} ,
\end{align}
For simplicity when plotting we use the parameters $\modsq{\alpha} = \modsq{\beta} =\frac12$ for the single rail encoding, as they are near-optimal and provide representative behaviour of all information measures considered here.
These are plotted together in Fig.~\ref{fig:ME-compare-single-dual-mixed-full}.

The conditional entropy is related to the mutual information via $S(\rho_A | \rho_R ) = S(\rho_A) - S(\rho_A ; \rho_R )$.
We calculate it for the quantum case $\ket{\psi}$ in the single rail encoding as
\begin{align}
  \label{eq:CEquantumsingle}
S\left(\rho_A|\rho_R\right) = & - \sum_{n=0}^\infty \left[ \frac{\modsq{\alpha}}{\cosh^2r} \tanh^{2n}r
                                        \lb{\tanh^2r \left(\frac{\modsq{\alpha}\cosh^2r + \modsq{\beta}(n+1)}
                                            {\modsq{\alpha}\sinh^2r + \modsq{\beta}n}\right)}\right. \nonumber \\
                                 & + \left. 2n \tanh^{2n}r \frac{\modsq{\beta}}{\cosh^2r} \left( \frac{n+1}{\cosh^2r}-\frac{n}{\sinh^2r}\right) \lb{\tanh r} \right. \nonumber
                                  + \frac{\modsq{\beta}}{\cosh^2r} \tanh^{2n}r \frac{n+1}{\cosh^2r} \lb{\frac{\modsq{\alpha}}{\cosh^2r}+\frac{\modsq{\beta}(n+1)}{\cosh^4r} } \nonumber \\
                                  & - \left. \frac{\modsq{\beta}n}{\sinh^2r\cosh^2r} \tanh^{2n}r\lb{\frac{\modsq{\alpha}}{\cosh^2r} + \frac{\modsq{\beta}n}{\sinh^2r\cosh^2r}}  \right].
\end{align}
The conditional entropy for the dual rail quantum case, using the parameter values $\modsq{\alpha} = \modsq{\beta} =\frac12$ is given by
\begin{align}
  \label{eq:CEquantumdual}
S\left(\rho_A|\rho_R\right)= - \sum_{p=0}^\infty \frac{\tanh^{2p}r}{\cosh^6r} \lb{\frac{p+2}{p+1}} \sum_{q=0}^p (q+1) .
\end{align}
\end{widetext}

\begin{figure}[b]
  \begin{center}
      \includegraphics[width=8.5cm]{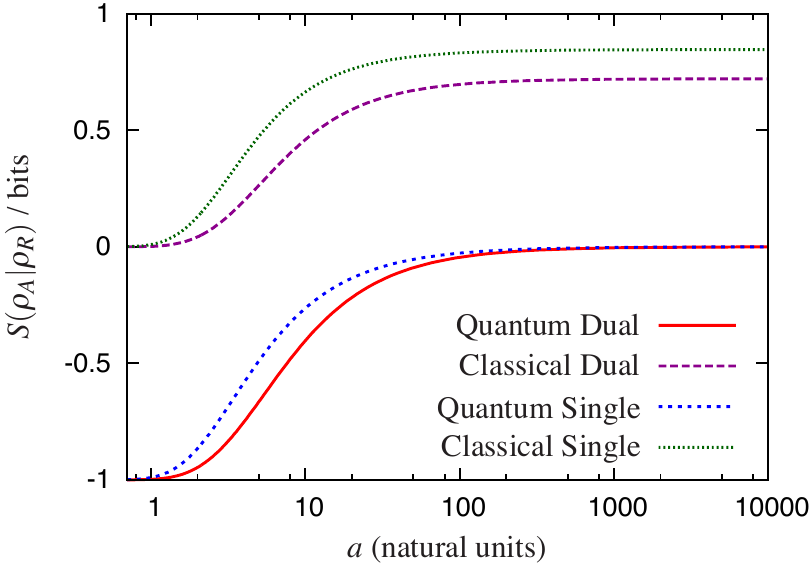}
  \caption[Conditional Entropy]{Conditional Entropy as a function of the acceleration $a$ when $\modsq{\alpha} = \modsq{\beta} =\frac12$.}
  \label{fig:conditionalentropy}
  \end{center}
\end{figure}

Using the same parameter values for the single rail case, we obtain the conditional entropy shown in Fig.~\ref{fig:conditionalentropy}. The graph starts at zero for both classical communication methods, showing that Rob fully learns Alice's state when there is no squeezing. We also find that as the squeezing increases, Rob needs more and more extra information to be able to fully specify Alice's state. For quantum communication, we see that the conditional entropy starts at $-1$ in the absence of acceleration ($a=0$). This means that Alice and Rob share entanglement, which allows them to share a qubit in the future using additional classical communication. The conditional entropy tends to zero when $a\to\infty$.
Note that we have not taken relativistic time dilation into account, which will affect the \emph{rate} of each communication channel in equal measure.

\begin{figure}[t]
  \begin{center}
      \includegraphics[width=8.5cm]{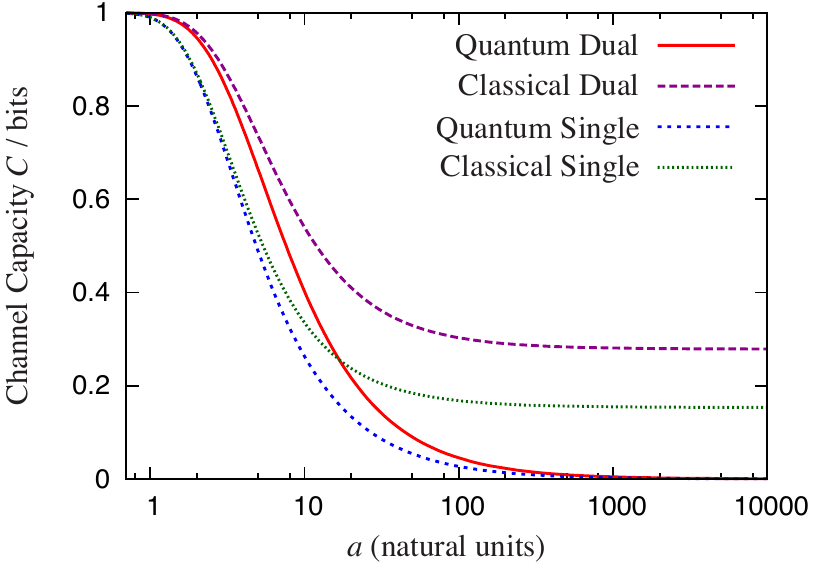}
  \caption[Channel Capacity]{The channel capacity $C$ as a function of the acceleration $a$ when $\modsq{\alpha} = \modsq{\beta} =\frac12$. The figure does not take into account relativistic time dilation, which reduces all communication \emph{rates} to zero equally. While the classical channel capacity reaches a finite value at the Schwarzschild horizon, the quantum coherent information (also measured in bits) drops to zero. Dual rail encoding outperforms single rail encoding in both classical and quantum communication.}
  \label{fig:compare-capacities}
  \end{center}
\end{figure}

Next, we compare the resulting channel capacities and coherent information. The classical channel capacity measures the amount of recoverable classical bits, while the quantum coherent information measures the entanglement that is generated using the state merging protocol. This quantum channel requires a classical channel of unlimited capacity. The comparison of all four channel capacities is shown in Fig.~\ref{fig:compare-capacities}.
When Rob is hovering far away from a black hole, there is little Unruh-Hawking noise ($a\simeq 0$), and all channel capacities are equal to one. Hovering closer to the black hole increases the noise and consequently reduces the channel capacity. Both classical channels tend towards finite values in the limit of large $a$, indicating that even when the communication rate drops to zero due to time dilation, each message that reaches Rob carries a finite amount of information.
This is explained by the fidelity between the logical bit states sent by Alice, which never reaches one when $a\to\infty$ (see
Fig.~\ref{fig:fidelity-of-zero-vs-one}).

On the other hand, the quantum coherent information falls off exponentially with $r$ as Rob hovers closer to the Schwarzschild horizon [$C \propto \exp(-\gamma r)$ for large $r$, with $\gamma\approx 2$ for the dual rail quantum channel].
Relating this back to the acceleration and distance from the horizon, we find that the quantum coherent information fully drops to zero for infinite acceleration, or at the horizon.
This is a fundamental difference between classical and quantum communication.
Classical communication relies on the occupation number, where, even at infinite acceleration, there is always a difference.
Quantum communication however relies on coherence, which gets completely washed out at infinite acceleration.

\section{Conclusion}\label{sec:conclusions}
\noindent
In conclusion, we studied communication channels between an inertial observer Alice and an accelerated observer Rob, using both single and dual rail encoding of a bosonic field. We found that the quantum coherent information tends to zero with increasing acceleration (i.e., approaching the horizon). The classical channel remained finite arbitrarily close to the Schwarzschild horizon, indicating that statistical correlations still exist between Alice and Rob even in the limit of infinite acceleration, whereas quantum correlations are fully removed. In both cases the channel degradation is due to Unruh-Hawking noise, and in both cases we ignored the time dilation that affects the rate of all communication channels equally.

\section*{Acknowledgments}
\noindent
The authors thank M. Wilde for valuable comments on the manuscript, and J. Dunningham and V. Palge for stimulating discussions. D. Hosler is funded by a University of Sheffield Studentship.

\end{document}